\documentclass[preprint,5p,twocolumn,superscriptaddress,color,showpacs]{elsarticle}

\usepackage{amsmath,amsfonts,amssymb,color,verbatim,wasysym}
\newcommand{\np}{NExP}

\begin{document}

\begin{frontmatter}

\title{No-exclaves percolation on random networks}

\author[inst1]{Byungjoon Min}
\ead{bmin@cbnu.ac.kr}
\author[inst2]{Eun-Kyu Park}
\author[inst2]{Sang-Hwan Gwak}
\author[inst2,inst3]{K.-I.\ Goh}
\ead{kgoh@korea.ac.kr}

\address[inst1]{Department of Physics, Chungbuk National University, Cheongju 28644, Korea
}
\address[inst2]{Department of Physics, Korea University, Seoul 02794, Korea
}
\address[inst3]{Department of Mathematics, University of California Los Angeles, Los Angeles, CA 90095, USA
}
\date{\today}

\begin{abstract}
No-exclaves percolation (NExP) is a nonlocal percolation process in which the components are formed not 
only by the connected occupied nodes but also by the agglomeration of empty nodes completely surrounded by the occupied nodes. 
It has been studied in low dimensions, displaying such novel phenomena as the discontinuous transition 
to complete percolation. However, its characteristics in complex networks are still unexplored. 
In this paper, we study the NExP on random networks by developing mean-field solutions using 
the generating function formalism. 
Our theory allows us to determine the size of the giant no-exclaves component as well as 
the percolation threshold, which are in excellent agreements with Monte Carlo simulations 
on random networks and some real-world networks. 
We show that on random networks NExP exhibits three phases and two transitions between them: 
the phases are characterized by the presence or absence of not only the giant NExP
component but also the giant unoccupied component, which is the giant connected component 
composed solely of unoccupied nodes. This work offers theoretical understanding on 
the anatomy of phase transitions in the NExP process. 
\end{abstract}


\end{frontmatter}

\section{Introduction}

Understanding percolation of complex networks has an impact on our ability to predict and control 
the behaviors of complex systems as it can provide essential information regarding the connectivity 
and robustness of complex systems~\cite{stauffer,error_attack,newman_book,dlee}. Classical percolation 
models are based on the notion of connectivity in which a pair of nodes is  considered a part of the 
same connected component if there exists at least one connected path between them~\cite{stauffer}. 
This model has provided a simple yet powerful theoretical framework across various phenomena, 
including epidemic spreading~\cite{newman02,pastor02}, traffic flows~\cite{transport}, 
the stability of power grid~\cite{power,yang,chilean}, and
the resilience of networks~\cite{error_attack,callaway,cohen2000}. There have also been numerous 
studies on the variants of standard percolation on networks that have modified the concept 
of connectivity such as $k$-core percolation~\cite{kcore}, bootstrap percolation~\cite{bootstrap}, 
$k$-selective percolation~\cite{kim_goh}, and mutual percolation~\cite{buldyrev,son,baxter12,min14}.

One of the most important applications of percolation theory in complex networks is the robustness of 
networks against random failures or attacks~\cite{error_attack,callaway,cohen2000,cohen2001,holme}.
The robustness of networked systems is often measured by how the size of the largest 
connected component responds when some nodes in networks are failed~\cite{error_attack,holme}.
The premise underlying the measurement is that it is essential for nodes to maintain their 
connectivity in order to function properly~\cite{error_attack,buldyrev}. 
Additionally, when nodes become non-functioning due to failure or detachment from the largest component
these nodes are assumed to be unable to recover permanently in most studies on the 
robustness of networks~\cite{error_attack,newman_book}. In reality, however, 
damaged nodes can often recover with communications to properly functioning nodes that 
surround the damaged nodes~\cite{min14,recovery,shekhman16,shang,wu2022}. For instance, temporary 
breakdown in real-world networks such as the brain and financial systems can recover
as long as the damage is localized.

The no-exclaves percolation ({\np}) is a model incorporating such a recovery rule that the 
component of failed nodes that is completely surrounded by active (unfailed) nodes 
recovers to be unfailed and merges into the surrounding unfailed components~\cite{nexp}. 
The failed component that is surrounded by unfailed nodes is referred to as the ``exclave'' 
which is the terminology in political geography and means a part of a district geographically 
isolated from the mainland by surrounding alien territories.
Our model does not allow for exclaves, hence the name ``no-exclaves percolation,'' or {\np} 
for short. After applying the no-exclaves rule for the recovery, we identify connected 
components and assess the robustness of networks using the concept of the giant {\np}
component. The no-exclaves rule in our model is closely 
related to the no-enclaves percolation on two dimensions offering a theoretical 
explanation for a motor-driven collapse of cytoskeletal systems~\cite{sheinman,gunnar,reply,hu2016}. 
Note that the no-enclaves rule does not readily apply to complex networks; 
the no-exclaves rule, however, is applicable to generic networks, while keeping 
the ingredient of nonlocality intact. Furthermore, the no-exclaves rule implies 
long distance communications between active nodes in some sense~\cite{castellano2020}.
From this perspective, it might have potential applicability to quantum communication 
and quantum network issues as well~\cite{meng,extended}.

It is known from numerical simulations that in low-dimensional systems 
the no-exclaves rule significantly 
affects the behavior of percolation transitions in spatial networks 
and Euclidean lattices~\cite{nexp}. 
However, there is still a lack of study for {\np} on networks as well as the 
analytical understanding of {\np} process. 
The main aim of the study is to make progress in filling this gap by developing the exact mean-field 
solutions to {\np} based on the generating 
function method in order to address the resilience of complex networks. 
We derive mean-field solutions to {\np} on random
networks, leading to an analytical insight to the effect of the no-exclaves rule and the anatomy of 
novel phase transitions displayed by the {\np} process.

The paper is organized as follows. In Sec.~II, we introduce the no-exclaves percolation model 
and explain how to apply no-exclave rules. We also introduce the concept
of the giant {\np} component as the order parameter. Next, we derive an analytical theory 
by using the generating function formalism, to compute the size of the giant {\np} component and 
the average {\np} component size in random networks (Sec.~III). 
Based on the theory, we identify and analyze the location of percolation transitions.
Finally, we apply our theory to random regular networks,
Erd\H{o}s-R\'enyi networks, and real-world networks from empirical data (Sec. IV).
The summary and discussions are presented in Sec.~V.

\section{No-Exclaves Percolation Model}

In this section, we present the no-exclaves percolation on networks, which
was originally studied for lattices~\cite{nexp}. The {\np} process is defined as follows.
Initially, all nodes are in an empty state in a given network. With the occupation 
probability $q$, we set each node to be occupied, randomly and independently. In terms of the network robustness, 
$1-q$ corresponds to failure probability. So far, the procedures 
are the same as the usual site percolation problem. Next, we identify the exclaves, 
which are the components composed of unoccupied nodes that are completely surrounded by the occupied nodes, except 
for the largest such one. 
Finally, the exclave gets merged into a bigger component comprising itself and all the neighboring  occupied components. 
Note that during this process multiple connected components, and even multiple exclaves, can merge 
into a single, larger component. We refer to the resulting combined components as the {\np} components.
We then measure the size $\Phi$ of the largest 
{\np} component as an order parameter and use occupation probability $q$ as a control 
parameter. The size $\Phi$ is equivalent to the probability that a randomly selected 
node belongs to the largest {\np} component.

Figure 1 shows an example of the no-exclaves percolation. Each node in a network is occupied 
with occupation probability $q$ [marked blue in Fig.~1(a)] or empty with probability $1-q$ 
[marked white in Fig.~1(a)]. Then, we identify the exclaves red-shaded in Fig.~1(b), the components 
of connected empty nodes that are completely surrounded by occupied nodes.
Practically, if all links leading out from the components composed of empty nodes connect 
to occupied nodes, the components  correspond to exclaves. 
Note that the ``giant'' connected component composed of empty nodes is not an exclave. 
Next, the exclaves merge with the neighboring occupied components, becoming 
``virtually''-occupied themselves as shown in Fig.~1(c). 
After agglomerating all the 
exclaves and surrounding occupied components, we identify the {\np} component, the connected 
component of both real- and virtually-occupied nodes [Fig.~1(d)]. 
The largest {\np} component in (d) is in general greater than the largest connected component in (a).

\begin{figure}
\includegraphics[width=\linewidth]{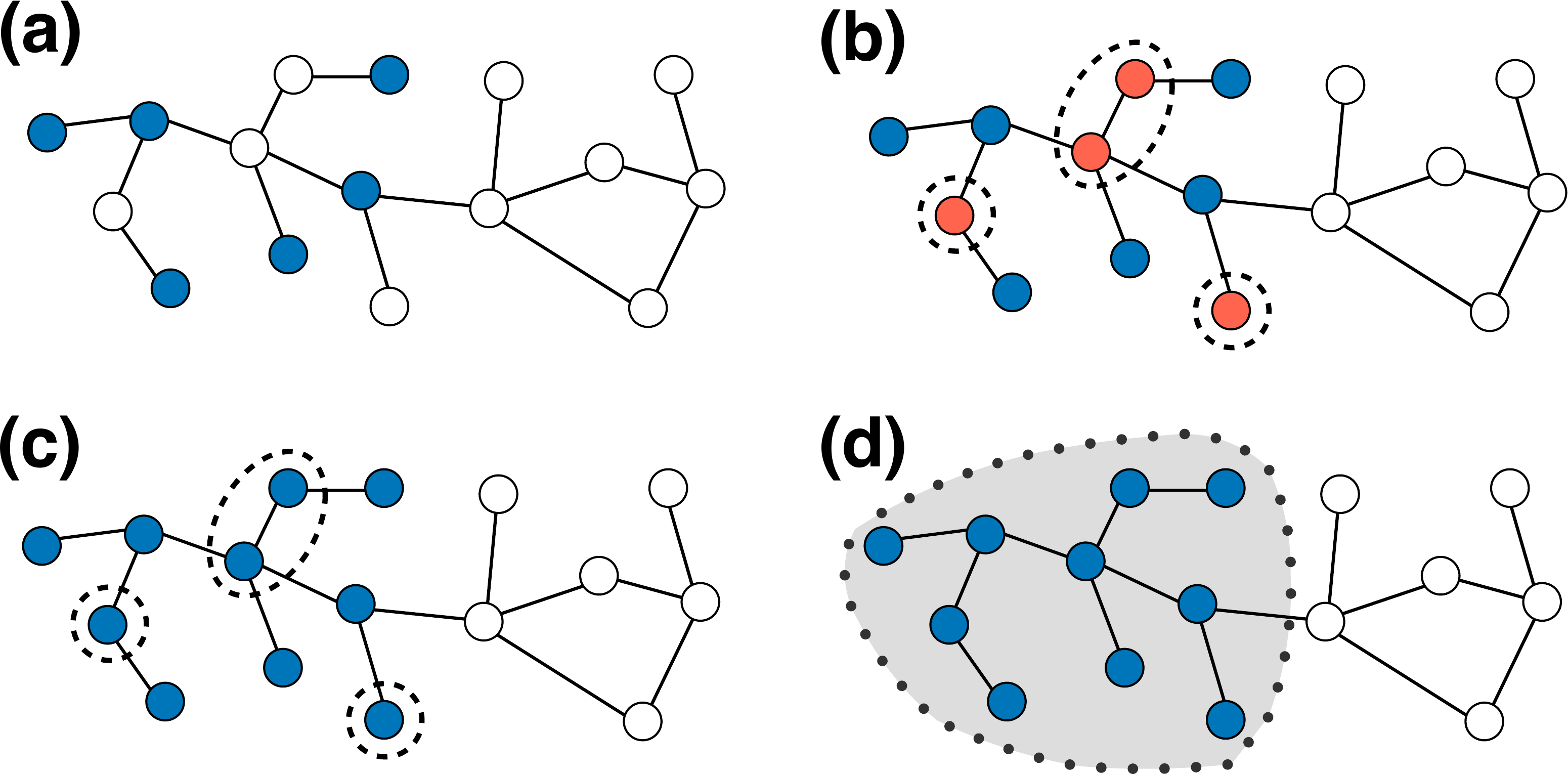}
\label{fig1}
\caption{
Schematic illustrations of {\np} on a network. (a) Each node in the network is either 
occupied (blue) with a probability $q$, or empty with a probability $1-q$. 
(b) Exclaves (red), components of empty nodes surrounded by occupied nodes, are identified. 
(c) The exclaves merge with surrounding components. 
(d) The largest {\np} component is identified, marked by the grey shaded area. 
}
\end{figure}

\section{Analytic Solution of NExP}

In this section, we derive the analytic solution of {\np} on random graphs using
the generating function formalism, applicable to locally-treelike networks in the thermodynamic limit. 
We observe that according to the no-exclaves rule, there can be three categories of components 
in {\np} on large networks: i) a giant {\np} component, ii) small (non-giant) {\np} components, and 
iii) a giant unoccupied component. 
A giant unoccupied component, which contains a finite fraction of nodes in the thermodynamic 
limit and extends to infinity, cannot be completely surrounded and thus is not an exclave. 
Note that we include isolated unoccupied components within the {\np} component 
as there is no need to distinguish between them for our purposes.

Based in this observation, our strategy for the analytical solutions of {\np} on networks is the following: 
We first identify the giant unoccupied component that should be excluded
when obtaining {\np} components by using a standard technique for the site percolation~\cite{newman}. 
The major difference is that we here consider a giant ``unoccupied'' component rather 
than occupied one that has been normally focused on. 
Then we derive self-consistency equations for the probability that a node arrived by following 
a randomly chosen link belongs to small {\np} component. 
The size of the giant {\np} component, denoted as $\Phi$, is determined by the complementary 
probability to the sum of the giant unoccupied component and the small {\np} components.

\subsection{The size of giant unoccupied component}

As noted above, we first calculate the size $S_0$ of a giant unoccupied component 
in order to obtain the size $\Phi$ of the giant {\np} component. 
The size $S_0$ can be obtained by following a conventional technique for 
the site percolation problem~\cite{newman}. 
It is important to note that since we calculated  the size of the giant unoccupied 
component, we determine the giant component among ``unoccupied'' nodes, rather 
than among occupied nodes as in ordinary site percolation problems.

Consider a random network with its degree distribution $P(k)$.
We define the degree and excess degree generating functions 
\begin{align}
{G}_0(x)&=\sum_{k=0}^{\infty} P(k) x^k, \\
{G}_1(x)&=\sum_{k=1}^{\infty} \frac{k P(k)}z x^{k-1},
\end{align}
where the excess degree stands for the degree for a node reached by following a randomly chosen link~\cite{newman}, 
and $z$ stands for the mean degree, $z=\langle k\rangle =\sum_k kP(k)$.
Let $R(s)$ be the probability that a randomly chosen node belongs to a small (non-giant) 
unoccupied component with size $s$. Here an unoccupied component stands for a connected 
component composed solely of unoccupied nodes. Then we define a generating function 
of $R(s)$ as ${H}_0(x)=\sum_{s=0}^{\infty} R(s) x^s$. 
We also define a generating function ${H}_1(x)$ that generates the size of a small 
unoccupied component reached by following a randomly chosen link.

On locally tree-like networks, the size of the small unoccupied component 
to which the node $i$ reached by following a randomly chosen link belongs is 
either zero if $i$ is occupied (with probability $q$) or 
the sum of the sizes of the unoccupied components to which $i$'s neighbors belong plus 
one (for $i$ itself) if $i$ is unoccupied (with probability $1-q$). 
Therefore, ${H}_1(x)$ and $H_0(x)$ satisfy the following self-consistency equations
\begin{align}
{H}_1(x) &=  q + (1-q) x {G}_1 ( {H}_1(x)), \label{eq:h} \\
{H}_0(x) &= q + (1-q) x {G}_0 ( {H}_1(x)).
\end{align}

We define $w$ as the probability for a node arrived at following a randomly 
chosen link does not belong to a giant unoccupied component. According to the definition
of $H_1(x)$, the value of $w$ is identical to $H_1(1)$ and given by Eq.~(\ref{eq:h}) 
\begin{align}
\label{eq:w}
w=q + (1-q) {G}_1 (w).
\end{align}
In addition, from the definition ${H}_0(x)$, we have the total probability
that a randomly chosen node belongs to a small unoccupied component as 
${H}_0(1) = \sum_{s=0}^{\infty} R(s)$.
Note that it is not necessarily normalized such that ${H}_0(1)=1$
because there may exist a giant unoccupied component that is not included in $R(s)$. 
The size of the giant unoccupied component, $S_0$, on a random network is equal to 
the probability that a randomly chosen node belongs to the giant unoccupied component. 
In conclusion, once we obtain $w$, the size $S_0$ of the giant unoccupied component 
can be obtained as 
\begin{align}
\label{eq:s0}
S_0 = (1-q) \left[ 1- G_0(w) \right].
\end{align}

\subsection{Self-consistency equations for {\np} component sizes}

In this section, we 
derive the self-consistency equations for {\np} component sizes.
Let $Q(\phi)$ be the probability that a randomly chosen node belongs to a finite (non-giant) 
{\np} component with size $\phi$ excluding both the giant {\np} component and the giant 
unoccupied component. Then we define the generating function
of $Q(\phi)$, represented  as $F_0(x)=\sum_{\phi=0}^{\infty} Q(\phi) x^\phi$. 
We also define the generating 
function ${F}_1(x)$ that generates the size of finite {\np} components for a node 
reached by a randomly chosen link. Note that $Q(\phi=1)$ may not be normalized to be unity since it 
excludes both the giant {\np} component and the giant unoccupied component.

We then derive self-consistency equations of ${F}_0(x)$ and ${F}_1(x)$ 
on a locally tree-like network. 
Assume that if we arrive at node $i$ which belongs to a small {\np} component 
by following a randomly chosen link, the size $\phi$ of the small component
including the node $i$ is one plus the sum of the sizes of the small 
components to which $i$'s neighbors belong. 
A key point to consider is that the probability of belonging to a size-$\phi$ component 
can vary depending on whether node $i$ is i) occupied or ii) unoccupied.

i) If node $i$ is occupied, a neighbor of node $i$ can be a part of either 
a small {\np} component or a giant unoccupied component. Note that if a neighbor 
belongs to a giant unoccupied component, 
it contributes by zero to the small component's size.
Then, the probability that a neighbor of an ``occupied'' 
node belongs to a small component with size $\phi$ is generated by $(1-w)+ {F}_1(x)$. 
There is an additional $(1-w)$ term with zeroth order in $x$ because it represents a probability for 
a giant unoccupied component, which is counted as if a zero-size {\np} component.

ii) On the other hand, if node $i$ is unoccupied, every neighbor of node $i$ cannot 
be a part of a giant unoccupied component. If it were, node $i$ would 
naturally belong to a part of the giant unoccupied component. Therefore, the probability 
that a neighbor of an ``unoccupied'' node belongs to a small component with size $\phi$ 
is generated by ${F}_1(x)$.

Summarizing these possibilities, we arrive at the following self-consistency 
equations for ${F}_0(x)$ and ${F}_1(x)$:
\begin{align}
F_1(x) &= q x  G_1 (1-w+F_1(x)) + (1-q) x G_1(F_1(x)),\label{eq:f} \\
F_0(x) &= q x  G_0 (1-w+F_1(x)) + (1-q) x G_0(F_1(x)).\label{eq:f0}
\end{align}

We define $u$ as the probability that a node by following a randomly 
chosen link belongs to a small {\np} component. This means 
it belongs to neither the giant {\np} component nor the giant unoccupied component. 
The probability $u$ is identical to $F_1(1)$ according to the definition of $F_1(x)$.
The value of $u$ can be obtained through numerical iteration of Eq.~(\ref{eq:f}), as given by
\begin{align}
u	= q {G}_1 ( 1-w+u ) + (1-q) {G}_1 ( u), \label{eq:u}
\end{align}
with $w$ from Eq.~(\ref{eq:w}).

\subsection{The size of giant {\np} component}

We are ready to obtain the giant {\np} component size~$\Phi$ on a random network which is identical to 
the probability that a randomly chosen node belongs to the giant {\np} component. 
The complementary probability of $\Phi$ is 
that a randomly chosen node either belongs to a small {\np} component, with probability $F_0(1)$,
or belongs to a giant unoccupied component, with probability $S_0$.
The probability $F_0(1)$ can be easily expressed as 
\begin{align}
{F}_0(1) = q  {G}_0 (1-w+u) + (1-q) {G}_0 (u),
\end{align}
according to Eq.~(\ref{eq:f0}).
Then, we finally arrive at the following expression for the size $\Phi$ of 
the giant {\np} component as
\begin{align}
\label{eq:phi}
\Phi &= 1- {F}_0(1) -S_0 \nonumber \\
 &= 1- q  {G}_0 (1-w+u) - (1-q) [1-{G}_0 (w)+G_0(u)], 
\end{align}
with $w$ from Eq.~(\ref{eq:w}) and $u$ from Eq.~(\ref{eq:u}).

In summary, our strategy for obtaining the size $\Phi$ of the giant {\np} component 
has been as follows. We first determine the probability $w$ that a node arrived by following 
a link does not belong to the giant unoccupied component by solving Eq.~(\ref{eq:w}), and subsequently 
the size $S_0$ of the giant unoccupied component by Eq.~(\ref{eq:s0}).
In addition, we obtain the probability $u$ that a node arrived by following a link 
belongs to a small {\np} component by Eq.~(\ref{eq:u}). Putting all together, we can
obtain the size $\Phi$ of the giant {\np} component by Eq.~(\ref{eq:phi}).

\subsection{The average {\np} size and percolation threshold}

We derive the average size $\chi$ of small {\np} components, excluding the 
giant {\np} component. When the giant {\np} component does not exist, a node reached by following 
a randomly chosen link must belong to either a small component with probability $u$ or a giant 
unoccupied component with probability $1-w$, resulting in the condition $1-w+u=1$. It simplifies to 
the condition $u=w$. With this condition, the average small {\np} component size can be expressed as
\begin{align}
\chi&=\left.\frac{d {F}_0(x)}{dx} \right\vert_{x=1} \nonumber \\
&=F_0(1)+ [q G_0'(1) + (1-q) G_0'(u) ] F_1'(1),
\end{align}
where $'$ denotes the derivative with respect to its argument $x$.
From Eq.~(\ref{eq:f}), we obtain 
\begin{align}
F_1'(1) =  \frac{u}{1- q G_1'(1) - (1-q) G_1'(u) }.
\end{align}
Since we consider the case when the giant {\np} component does not exist, 
the average component size is given by 
\begin{align}
\chi=F_0(1)+ \frac{ u[q G_0'(1) + (1-q) G_0'(u) ] }{[1- q G_1'(1) - (1-q) G_1'(u)] }.
\end{align}

The percolation threshold $q_c$ where a giant {\np} component first appears is 
located at the point where the average component size diverges, satisfying
\begin{align}
q_c G_1'(1) + (1-q_c) G_1'(u) =1.
\label{eq:pc}
\end{align}
The percolation threshold $q_c$ of the giant {\np} component is in general not amenable 
to compact expression in terms of $z$ or other moments of $P(k)$ because $w$ and $u$ also 
contain $q$, yet can be computed numerically. Beyond this threshold $q_c$, the size of 
the giant {\np} component continues to grow as more occupied nodes are added.

The {\np} threshold $q_c$ is smaller than the threshold of the random site percolation 
$q_c^{RSP}$, which is given by $q_c^{RSP} G_1'(1)=1$.
The reduction of the threshold is caused by the no-exclaves rule for recovery,
which leads to the merging of small and separated occupied components. 
This contribution is reflected in the second term of Eq.~(\ref{eq:pc}).
In the region $q_c<q<q_c^{RSP}$, multiple separated occupied components merge 
into a giant {\np} component during the recovery of exclaves;
An example of the merging is depicted in Fig.~1(c).

In {\np}, there is another transition where a giant unoccupied component disappears 
as $q$ increases. In a scenario where a giant unoccupied component vanishes, it becomes possible that the giant 
{\np} component would span the entire nodes within the giant connected component of the underlying undisturbed ($q=1$) network. We refer the point of this transition as the ``complete'' percolation 
point, denoted as $q^*$. The complete percolation point $q^*$ corresponds to the location
in which the size $\Phi$ of the giant {\np} component becomes unity if all nodes of
the underlying network belong to a single connected component.
The location of $q^*$ can be identified by the divergence of the average size $\langle s \rangle$
of unoccupied components. The condition for $q^*$ is given from Eq.~(\ref{eq:h}) as
\begin{align}
G_1'(1) = \frac{1}{1-q^*}.
\end{align}
Since $G_1'(1) = \frac{ \langle k^2 \rangle - \langle k \rangle}{\langle k \rangle}$ 
for random networks, $q^*$ can be  expressed compactly as 
\begin{align}
q^* &= 1- \frac{\langle k \rangle}{\langle k^2 \rangle - \langle k \rangle} = \frac{\langle k^2 \rangle - 2\langle k \rangle}{\langle k^2 \rangle - \langle k \rangle},
\label{eq:pstar}
\end{align}
which is directly related to the well-known result for the percolation on random networks~\cite{callaway,cohen2000,molloy1998}.

The transitions can be interpreted by the stability analysis of the solutions 
of Eqs.~(\ref{eq:w}) and (\ref{eq:u}). The complete percolation point $q^*$ corresponds
to the point where the trivial fixed point of $w$ in Eq.~(\ref{eq:w}) becomes unstable.
The trivial solution of Eq.~(\ref{eq:w}) is $w=1$ since $G_1(1)$ is normalized to be $1$. 
The solution $w=1$ corresponds to no giant unoccupied component considering the meaning 
of $w$. It means that the giant unoccupied component disappears at $q^*$.
On the other hand, the {\np} threshold $q_c$ can be obtained by the stability analysis
of Eq.~(\ref{eq:u}). The trivial solution of Eq.~(\ref{eq:u}) is $u=w$ because Eq.~(\ref{eq:u})
reduces into Eq.~(\ref{eq:w}) when $u=w$, or equivalently $1-w+u=1$. The solution $u=w$
represents no giant {\np} component and it becomes unstable at $q_c$. Therefore 
another solution becomes stable, leading to the emergence of the giant {\np}
component.

These two transition points $q_c$ and $q^*$ do not in general coincide on random networks. 
The percolation transition $q_c$ is generally smaller than the complete point $q^*$. 
Therefore, there can be three phases in {\np}, depending on the existence and extinction of 
the giant {\np} component and the giant unoccupied component: We have phases in which 
i) the giant unoccupied component exists but giant {\np} 
component does not when $q<q_c$; ii) both the giant {\np} component and the giant unoccupied 
component coexist when $q_c<q<q^*$; iii) the giant {\np} component exists but giant 
unoccupied component does not when $q>q^*$. 
The multiple percolation transitions in network {\np} are in stark contrast to the 2D {\np} in which 
the two occur at the same point \cite{nexp}. Our solutions suggest that the discontinuous percolation 
transition into complete percolation displayed by 2D {\np} can be resolved into two transitions combined into one.

\subsection{Critical exponents of {\np}}

We examine the critical exponents of {\np} in the vicinity of the {\np} threshold $q_c$
by using generating function method \cite{cohen2002}.
We first determine the order parameter critical exponent $\beta$, indicating that
$\Phi \sim (q-q_c)^{\beta}$.
The singular behavior of $\Phi$ stems from the self-consistency 
equations, Eqs.~(\ref{eq:w}) and~(\ref{eq:u}). 
We examine Eq.~(\ref{eq:w}) by rewriting with $w=w_c+\epsilon_w$ and $q=q_c+\delta$. 
Then, Eq.~(\ref{eq:w}) becomes
$w_c+ \epsilon_w  = (q_c +\delta) +(1-q_c - \delta) G_1(w_c+\epsilon_w)$.
Keeping only the leading order of $\epsilon_w$ for a small $\epsilon_w$, 
Eq.~(\ref{eq:w}) implies 
\begin{align}
\epsilon_w  &\sim \frac{1-G_1(w_c)}{q_c G_1'(1)} \delta.
\label{eq:ew}
\end{align}
Similarly, we expand Eq.~(\ref{eq:u}) for $q=q_c+\delta$, $w=w_c+\epsilon_w$.
and $u=u_c-\epsilon_u$ and obtain the relation 
\begin{align}
\epsilon_u \sim \frac{2 [G_1'(1)-G_1'(u_c)]}{q_c G_1''(1) +(1-q_c)G_1''(u_c)}\delta.
\end{align}
Finally, the size of the giant {\np} component grows near the 
{\np} threshold as:
\begin{align}
\Phi \sim c_w \epsilon_w + c_u \epsilon_u \sim (q-q_c)^{\beta},
\end{align}
leading to $\beta=1$ which agrees with the mean-field percolation result.

By employing the ansatz $Q(\phi) \sim \phi^{1-\tau}$ at $q=q_c$,
we can derive the exponent $\tau$.  
When we expand Eq.~(\ref{eq:f}) at $q=q_c$ for $x=1-\epsilon$
and $F_1(x)=u_c-\Delta$, the lowest order yields
\begin{align}
\epsilon \sim \frac{q_c G_1''(1)+(1-q_c) G_1''(u_c)}{2 u_c} \Delta^2.
\end{align}
At the {\np} threshold $q_c$, the average size of $\phi$
diverges, leading to the fact $2 < \tau \le 3$.
Then the generating function $F_1(x)$ should satisfy
$u_c-F_1(1-\epsilon) \sim \epsilon^{\tau-2}$.
Given that $\Delta \sim \epsilon^{1/2}$, 
we arrives at $\tau=5/2$, which corresponds to the same value of 
the mean-field percolation class.

Other critical exponents are determined by scaling relations \cite{stauffer,cohen2002}, 
and thus we suggest that the {\np} belongs to the mean-field universality class, 
based on the fact that $\beta=1$ and $\tau=5/2$.
In the case of scale-free networks, the critical exponents would differ from 
the mean-field values due to the dependency on degree exponents. 
However, we have not considered the effects of degree heterogeneity of 
scale-free networks in this study.

\section{Specific Results}

\subsection{On Random Regular Networks}

To gain insight on the solutions of {\np}, we derive explicit expressions 
of our theory applied to random regular (RR) networks with degree $z$
and a uniform occupation probability $q$. 
The degree distribution of RR networks is a Kronecker delta distribution, $P(k) = \delta_{k,z}$.  
In RR networks, the probability that a node reached by following a link belong to 
a finite unoccupied component $w$ and belong to a finite {\np} component $u$
are given by the coupled self-consistency equations as
\begin{align}
w & = q + (1 - q) w^{z-1},  \\ 
u & = q (1-w+u)^{z-1} + (1-q) u^{z-1}.
\end{align}

For an explicit example of $z=3$, the solutions of $w$ are $1$ and $q/(1-q)$, which correspond 
to the phase without a giant unoccupied component and with a giant unoccupied component, respectively.
The trivial fixed point $w=1$ corresponding to no giant unoccupied component 
becomes unstable when $q<q^*$ where $q^*$ is the complete percolation point.
The value of complete percolation point between the two phases is given by
\begin{align}
q^* = \frac{z-2}{z-1}=\frac{1}{2}.
\end{align}
Depending on $q$, the value of $w$ is explicitly given by 
\begin{align}
w=\left\{
	\begin{array}{ll} 
	\frac{q}{1-q},\quad  & q<q^*,  \\ 
	1, \quad  & q>q^*.   
	\end{array}\right. 
\end{align}
While there is the giant unoccupied component when $q<q^*$, 
the giant unoccupied component disappears when $q>q^*$.

When $w=1$ for $q>q^*$, $u$ becomes $0$. If $q<q^*$ indicating that 
$w=q/(1-q)$, the possible solutions of $u$ are either $u=q/(1-q)$ and $(1-2q)^2/(1-q)$.
The solution $u=w=q/(1-q)$ corresponds to the case when there is no giant {\np} component. 
If $u=(1-2q)^2/(1-q)$, $1-w+u$ becomes less than unity meaning that there is a
giant {\np} component. 
The {\np} threshold between two phases is given by
\begin{align}
w(q_c) &= \frac{1}{z}=\frac{1}{3}, 
\end{align}
which leads to
\begin{align}
q_c &= \frac{1}{4}.
\end{align}
Depending on $q$, the value of $u$ is given by 
\begin{align}
u=\left\{
	\begin{array}{ll} 
	\frac{q}{1-q},\quad  & q<q_c,  \\ 
	\frac{(1-2q)^2}{1-q},\quad  & q_c<q<q^*,  \\ 
  0, \quad  & q>q^*.
	\end{array}\right. 
\end{align}

Therefore we have three regimes of no-exclaves percolation in RR networks:
i) giant {\np} component does not exist but the giant unoccupied component
exists ($q<q_c$); ii) both the giant {\np} component and the giant unoccupied component 
coexist ($q_c<q<q^*$); and iii) only the giant {\np} component whose size is unity 
exists ($q>q^*$). Since we can express $u$ and $w$ in closed forms for RR networks 
with $z=3$, we can explicitly obtain the expression for the giant {\np}
component size $\Phi$ as: 
\begin{align}
\Phi=\left\{\begin{array}{ll}
	0, \quad & q<q_c~, \\
	\frac{(4q-1)(8q^3-6q^2+1)}{1-q},\quad  & q_c <q< q^*~, \\ 
	1,\quad  & q>q^*~. 
\end{array}\right. 
\end{align}
As shown in Fig. 2, the probabilities $w$ and $u$, and the giant {\np} component size $\Phi$ as a function of the occupation probability $q$ 
show perfect agreements with the Monte Carlo simulation results.

\begin{figure}[t]
\centering
\includegraphics[width=\linewidth]{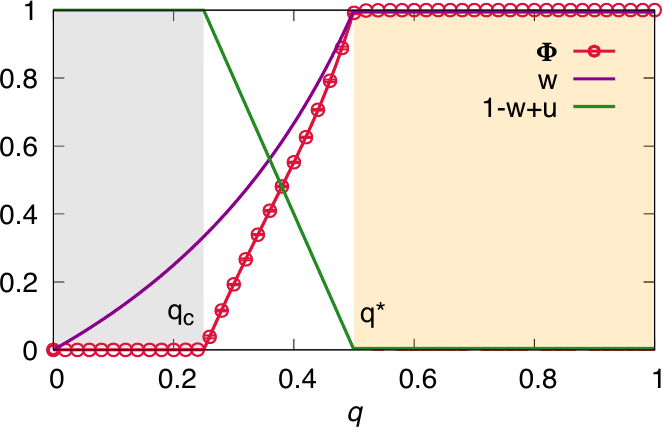}
\caption{
Analytic predictions of $\Phi$, $w$, and $1-w+u$ with respect to $q$ 
on RR networks with $z=3$ and $N=10^6$ are shown. Theoretical results (lines) and 
simulation results (points) are shown together.
Error-bars denote the standard deviation from $10^6$ independent runs. 
}
\end{figure}

\subsection{On Erd\H{o}s-R\'enyi Networks}

We consider the NExP model on Erd\H{o}s-R\'enyi (ER) networks, 
which have a Poisson degree distribution $P(k) = e^{-z}z^k/k!$ in the large $N$ limit, where $z$ is the average 
degree. The generating functions for ER networks are given by $G_0(x) = e^{z(x-1)}$ and  $G_1(x) = e^{z(x-1)}$. 
This leads us to the coupled equations for the variables $w$ and $u$, given by 
\begin{align}
w & = q + (1 - q)e^{z(w-1)}~, \label{eq:erw} \\
u & = q e^{z(u-w)} + (1-q) e^{z(u-1)}.
\label{eq:eru}
\end{align}
The complete percolation point and the percolation threshold for {\np} can be determined 
by Eqs.~(\ref{eq:pstar}) and (\ref{eq:pc}), respectively. These conditions show the instability 
of the trivial solutions, $w=1$ for the complete percolation point and $u=w$ for the 
percolation threshold, respectively. For ER networks, the specific values for these thresholds are given by
\begin{align}
q^* = 1-\frac{1}{z} \quad \textrm{and}\quad
w(q_c) =  \frac{1}{z}.
\label{eq:pcer}
\end{align}

We have three regimes in the {\np} model, similar to RR networks. 
i) When $q<q_c$, there is no giant {\np} component but the giant unoccupied component exists. 
ii) When $q_c<q<q^*$, both the giant {\np} component and the giant unoccupied component coexist. 
iii) When $q>q^*$, only the giant {\np} component exists. Furthermore, when $q>q_c$, 
the giant {\np} component size $\Phi$ can be expressed as
\begin{align}
\Phi &= 1-q e^{z(u-w)}-(1-q) [1- e^{z(w-1)} + e^{z(u-1)}],
\end{align}
where $w$ and $u$ can be  obtained by Eqs.~(\ref{eq:erw}, \ref{eq:eru}).

\begin{figure}[t]
\centering
\includegraphics[width=\linewidth]{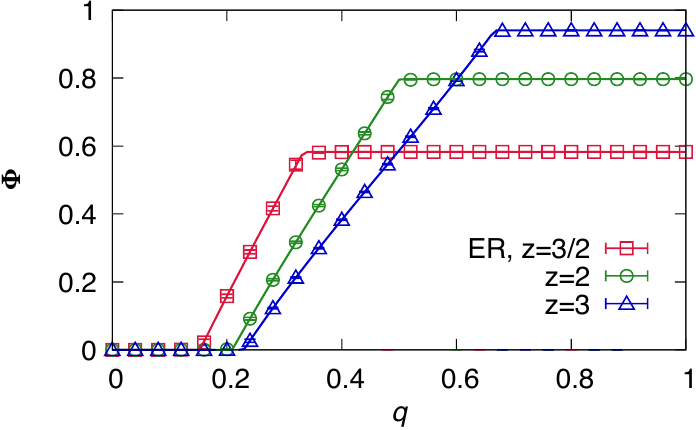}
\caption{
The giant {\np} component size $\Phi$ as a function of the occupation 
probability $q$ on ER networks with mean degree $z = 3/2, 2$, and $3$ and $N=10^6$. 
Numerical results (points) averaged over $10^6$ independent runs 
and analytic curves (lines) are shown together. 
}
\label{fig:er}
\end{figure}

We compare the theoretical results derived using the generating function method 
with the Monte Carlo simulations on ER networks of average degree $z=3/2, 2, 3$ 
and network size  $N=10^6$. Figure \ref{fig:er} shows the giant  
{\np} component size $\Phi$ obtained from our theory (lines) and from the numerical simulations 
(points) as a function of $q$. We observe excellent agreements between our theoretical 
predictions and the simulation results. 
Furthermore, we validate the locations 
of the percolation transition, $q_c$, and the complete 
transition point, $q^*$, which were predicted from Eq.~(\ref{eq:pcer}).

\subsection{On real-world networks}

We test our theory for NExP on top of real-world 
networks built from empirical data. We use the two datasets:
the Internet of an autonomous system level compiled by CAIDA \cite{caida} and
the United States airport network obtained from OpenFlights \cite{flights}.
The former represents the router network of Internet reconstructed from the packet flows between neighboring peers at the autonomous system level, collected on November 5, 2007. 
The U.S.\ airport network represents the connections from the origin
to destination airports in the year 2010. 
For the sake of simplicity, we ignore directionality
and weights of the links from the data when we construct the networks. 
The router network contains $N=26,464$ nodes with the mean degree $z=3.26$ and the airport network has $1,573$ nodes with $z=30.39$.

The theoretical predictions and numerical simulation results for the giant {\np} component size $\Phi$ 
are shown in Fig.~4. We confirm that our theory generates a reliable prediction 
in the size of the giant {\np} component for both the Internet [Fig.~4(a)] and 
the airport network [Fig.~4(b)]. We also measure the size of the 
giant connected component for random site percolation (denoted in short as RSP)
for comparison \cite{callaway,cohen2000}. We found that $\Phi$ for {\np} is higher, compared to that for the standard node removal scenarios. It concludes that real-world networked 
systems can be more robust than expected when considering the recovery of exclaves. 
It is noteworthy that the shape of $\Phi$-curve for the real-world networks is different 
from those of Fig.~3: It does not display the two-transition character clearly. This is due 
mainly to the fact that both the real-world networks are ``scale-free,'' having broad distribution 
of degrees, driving $q_c$ towards zero and $q^*$ towards unity. Finite-size effects can also be 
attributed for the smearing of the transitions in real-world networks. 
Even so, our theory works remarkably well on these real-world networks.

\begin{figure}[t]
\centering
\includegraphics[width=\linewidth]{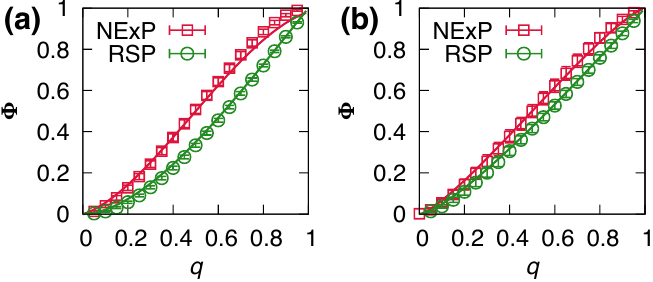}
\caption{
The size $\Phi$ of the giant {\np} component of two real-world networks:
(a) the Internet's router network at the autonomous system level and 
(b) the U.S.\ airport network.
Monte-Carlo simulation results (symbols) averaged over $10^4$ independent runs 
and analytic curves (lines) are shown together. 
}
\label{fig:real}
\end{figure}

\section{Summary and discussion}

In this study, we have focused on the no-exclaves percolation ({\np}) process 
on random networks and formulated 
a generating function approach to derive analytical solutions for this model. 
We identify the percolation threshold and the complete transition point for the emergence 
of the giant {\np} component and the disappearance of the giant unoccupied component, respectively. 
In addition, we apply the theory to random regular graphs, Erd\H{o}s-R\'enyi graphs,
and real-world networks, and confirm great agreements between our theoretical predictions 
and Monte Carlo simulations. 

From theoretical perspective, our solution offers theoretical understanding for the multiple phase 
transition structure of {\np} on random networks. It shows how the two-threshold behavior associated 
with the emergence of the giant {\np} component and the extinction of the giant unoccupied component 
emerge as the occupation probability $q$ increases in network {\np}. It also suggests that the single 
discontinuous percolation in 2D {\np}~\cite{nexp} may be resolved into two transitions converging 
at the same point.

From the perspective of network robustness, our study implies that the recovery processes 
according to the no-exclaves rule can effectively mitigate the impact of random failures of nodes.
When we remove nodes with failure probability $1-q$ as random failure problems~\cite{error_attack,cohen2000}, 
the giant ``occupied'' component disappears at $q^*$. However, by introducing
the no-exclaves rule, we can maintain the giant {\np} component at the point $q_c$,
which is in general less than $q^*$. In addition, we also found that 
the random failure above the complete percolation point $q^*$ does not make any effect 
in the no-exclaves percolation. In this sense, our study suggests
an effective way to improve the network robustness with 
the recovery of failed exclaves.

\section*{Acknowledgments} 
This work was supported in part by the National Research Foundation of Korea (NRF) grants 
funded by the Korea government (MSIT) (No.\ 2020R1A2C2003669 (K-IG) and No.\ 2020R1I1A3068803 (BM)).

\end{document}